\documentstyle[sprocl,twoside,amssymb]{article}
\def\citebk#1{\hspace{0.9mm}\raisebox{-1.85mm}[0mm][0mm]
  {\Large\cite{#1}}\hspace{-0.1mm}}

\def\citebkcap#1{\hspace{0.8mm}\raisebox{-1.5mm}[0mm][0mm]
  {\large\cite{#1}}\hspace{-0.2mm}}

\begin{document}

\newcommand{\Si}{\Sigma}
\newcommand{\tr}{{\rm tr}}
\newcommand{\ad}{{\rm ad}}
\newcommand{\Ad}{{\rm Ad}}
\newcommand{\ti}[1]{\tilde{#1}}
\newcommand{\om}{\omega}
\newcommand{\Om}{\Omega}
\newcommand{\de}{\delta}
\newcommand{\al}{\alpha}
\newcommand{\te}{\theta}
\newcommand{\vth}{\vartheta}
\newcommand{\be}{\beta}
\newcommand{\la}{\lambda}
\newcommand{\La}{\Lambda}
\newcommand{\D}{\Delta}
\newcommand{\ve}{\varepsilon}
\newcommand{\ep}{\epsilon}
\newcommand{\vf}{\varphi}
\newcommand{\G}{\Gamma}
\newcommand{\ka}{\kappa}
\newcommand{\ip}{\hat{\upsilon}}
\newcommand{\Ip}{\hat{\Upsilon}}
\newcommand{\ga}{\gamma}
\newcommand{\ze}{\zeta}
\newcommand{\si}{\sigma}
\def\bfa{{\bf a}}
\def\bfb{{\bf b}}
\def\bfc{{\bf c}}
\def\bfd{{\bf d}}
\def\bfe{{\bf e}}
\def\bfm{{\bf m}}
\def\bfn{{\bf n}}
\def\bfp{{\bf p}}
\def\bfu{{\bf u}}
\def\bfv{{\bf v}}
\def\bft{{\bf t}}
\def\bfx{{\bf x}}
\def\bfg{{\bf g}}
\def\bfS{{\bf S}}
\def\bfJ{{\bf J}}
\def\bfxi{{\bf \xi}}
\newcommand{\ran}{\rangle}
\newcommand{\lan}{\langle}
\newcommand{\lst}[1]{{\la #1|}}
\newcommand{\st}[1]{{|#1\ra}}
\newcommand{\oper}[2]{{\st{#1}\lst{#2}}}
\newcommand{\li}{\lim_{n\rightarrow \infty}}
\newcommand{\mat}[4]{\left(\begin{array}{cc}{#1}&{#2}\\{#3}&{#4}
\end{array}\right)}
\newcommand{\thmat}[9]{\left(
\begin{array}{ccc}{#1}&{#2}&{#3}\\{#4}&{#5}&{#6}\\
{#7}&{#8}&{#9}
\end{array}\right)}
\newcommand{\beq}[1]{\begin{equation}\label{#1}}
\newcommand{\eq}{\end{equation}}
\newcommand{\beqn}[1]{\begin{eqnarray}\label{#1}}
\newcommand{\eqn}{\end{eqnarray}}
\newcommand{\p}{\partial}
\newcommand{\di}{{\rm diag}}
\newcommand{\oh}{\frac{1}{2}}
\newcommand{\su}{{\bf su_2}}
\newcommand{\uo}{{\bf u_1}}
\def\GLN{{\rm GL}(N, {\mathbb C})}
\def\sln{{\rm sl}(N, {\mathbb C})}
\def\SLN{{\rm SL}(N, {\mathbb C})}
\def\Sl3{{\rm SL}(3, {\mathbb C})}
\def\SL2{{\rm SL}(2, {\mathbb C})}
\def\GL2{{\rm GL}(2, {\mathbb C})}
\def\sl2{{\rm sl}(2, {\mathbb C})}
\def\gl2{{\rm gl}(2, {\mathbb C})}
\newcommand{\gln}{{\rm gl}(N, {\mathbb C})}
\newcommand{\PSL}{{\rm PSL}_2( {\mathbb Z})}
\def\f1#1{\frac{1}{#1}}
\def\lb{\lfloor}
\def\rb{\rfloor}
\def\sn{{\rm sn}}
\def\cn{{\rm cn}}
\def\dn{{\rm dn}}
\def\bfS{{\bf S}}
\def\bfT{{\bf T}}
\def\we{\wedge}
\newcommand{\rar}{\rightarrow}
\newcommand{\upar}{\uparrow}
\newcommand{\sm}{\setminus}
\newcommand{\ms}{\mapsto}
\newcommand{\bp}{\bar{\partial}}
\newcommand{\bz}{\bar{z}}
\newcommand{\bA}{\bar{A}}
\newcommand{\bL}{\bar{L}}
\newcommand{\vtb}{\theta_{10}}
\newcommand{\vtc}{\theta_{00}}
\newcommand{\vtd}{\theta_{01}}
\newcommand{\sect}[1]{\setcounter{equation}{0}\section{#1}}
\renewcommand{\theequation}{\thesection.\arabic{equation}}
\newtheorem{defi}{Definition}
\newtheorem{rem}{Remark}[section]
\sloppy

\title{LIE ALGEBROIDS AS GAUGE SYMMETRIES IN TOPOLOGICAL
FIELD THEORIES}

\author{M.A. OLSHANETSKY}
\address{Max Planck Institute of Mathematics,     
Vivatsgasse 7, D-53111 Bonn, Germany\\and\\
Institute of Theoretical and Experimental Physics, 117259, 
Moscow, Russia}

\maketitle

\vspace{0.4cm}

\abstracts{
The Lie algebroids are  generalization of the Lie algebras.
They arise, in particular, as a mathematical tool in investigations 
of dynamical systems with the first class constraints. Here we consider 
canonical symmetries of Hamiltonian systems generated by a
 special class of Lie algebroids.
 The ``coordinate part'' of the Hamiltonian phase space is the 
Poisson manifold $M$ and the Lie algebroid brackets are defined by  means
of the Poisson bivector. The Lie algebroid action defined on $M$ can be lifted to
the phase space. 
The main observation is that the classical BRST operator
has the same form  as in the case of the Lie groups action. 
Two examples are analyzed. In the first, $M$ is the space of $SL(3,C)$-opers on
 Riemann curves with the Adler-Gelfand-Dikii brackets. 
The corresponding Hamiltonian
system is the $W_3$-gravity. Its  phase space is the base of the algebroid bundle. 
The sections of the bundle are the second order differential operators on Riemann
 curves. 
They are the gauge symmetries of the theory. The moduli space of $W_3$ geometry
of Riemann  curves is the symplectic quotient with respect to their action. 
It is demonstrated
that the nonlinear brackets and the second order differential operators
arise from the canonical brackets and the standard gauge transformations
in the Chern-Simons field theory, as a result of the partial gauge fixing.
The second example is $M=C^4$ endowed with the Sklyanin brackets.
The symplectic reduction with respect to the algebroid action leads to
a generalization of the rational Calogero-Moser model. As in the previous
example the Sklyanin brackets can be derived from a ``free theory.'' 
In this case it is a ``relativistic deformation'' of the 
$SL(2,C)$ Higgs bundle over an elliptic curve.
}


\vspace{0.4cm}

\tableofcontents

\newpage

\section{Introduction}
\setcounter{equation}{0}

In this paper I analyze the special first class constraints in the classical
Hamiltonian systems. This subject was one of the main interests of Misha
Marinov. I remember very interesting discussions with him when he prepared with 
Misha Terentev their paper devoted to dynamics on group manifolds.\cite{MM}
They used the path integral for the quantization and
the Lie group symmetries played essential role in their construction.
The aim of this paper is going beyond the group symmetry, though we
consider only the classical theory.
I hope that Misha would have read to the present paper with interest.\footnote{~Sections 
2-4 of this paper is an updated version of part of Ref.~\protect\citebkcap{LO}.}

In fact, the group symmetries  by no means exhaust 
 all interesting symmetries in classical and quantum 
Hamiltonian systems.
Generic first class constraints  generate transformations that generalize
the Lie group action. For example, the structure constants can depend
on the points of the phase space. A generalization of this situation leads
to the notion of the Lie algebroid and Lie groupoid. Batalin is likely to
be the first who explicitly used this construction in gauge theories.\cite{Ba}
He called these transformations the quasigroups.

The BRST approach allows to work with arbitrary forms of constraints.
In general situation the BRST operator has an infinite rank
(the degree of the ghost momenta), while in
the Lie group case it has rank one  or less.\cite{HT} Nevertheless,
it is possible to modify the Lie group symmetries in a such way that
 the BRST operator has the same form as for the Lie groups.
We consider one of these cases. 

To construct these symmetries we carry out the following steps:\\
1. Consider a Poisson manifold $M$. The transformations of $M$ 
 depend on the Poisson structure.
The local version of them is a particular type of Lie algebroids ${\cal A}_M$.
The space $M$ plays the role of the ``coordinate subspace'' and the algebroid
action is the transformations of the coordinates. \\
2. To come to the field theory we introduce
 the space of maps \mbox{${\bf M}=D\to M,~D=\{|z|\le 1\}$}.  The
algebroid ${\cal A}_{\bf M}$ is the analogue of the loop algebra. 
It can be central extended $\hat{\cal A}_{\bf M}$ by a one-cocycle
as in the loop algebra case.\\
 3. The transformations coming from the Lie algebroid $\hat{\cal A}_{\bf M}$
can be lifted to the  phase space ${\cal R}$. In addition to the coordinate
space ${\bf M}$ the phase space
 ${\cal R}$ contains the conjugate variables. It is 
endowed with the  canonical symplectic structure. The space ${\cal R}$
along with its symmetries defines the so-called Poisson sigma-model.\cite{I,SS}\\
4. The infinitesimal form of the symmetry transformations we call
the Hamiltonian algebroid ${\cal A}^H_{\cal R}$ related to the Lie algebroid 
${\cal A}_M$.
The Hamiltonian algebroid is a generalization of the Lie algebra of symplectic
vector fields with respect to the symplectic structure on ${\cal R}$.
The special feature of these systems is that the BRST operator has 
the same structure as for the transformations performed by Lie algebras.

We present two examples of topological field theories
with these symmetries. The first example is the
$W_3$-gravity\,\cite{P,BFK,GLM} and related to this theory
 generalized deformations
of complex structures of Riemann curves by the second order differential
operators. This theory is a generalization
of $2+1$-gravity ($W_2$-gravity),\cite{Ca}
where the space component has a topology of
a Riemann curve $\Si_{g}$ of genus $g$.
The Lie algebra symmetries in $W_2$-gravity is the algebra of
smooth vector fields on $\Si_{g}$. After killing the gauge degrees of freedom
one comes  to the moduli space of projective
 structures on $\Si_{g}$. These structures can be described by the BRST method
which is straightforward in this case. The case of $W_N$-gravity $(N>2)$ is more 
subtle.
The main reason
 is that the gauge symmetries do not generate the Lie group action.
This property of $W_N$-gravity was well known.\cite{P}

We consider here in detail the $W_3$ case.
The infinitesimal symmetries are carried out by the second order differential
operators on $\Si_{g}$ without constant terms. 
The role of the Poisson manifold $M$ is played by the $M_3=\Sl3$-opers\,\cite{Te,BD}
 endowed with the Adler-Gelfand-Dikii bivector.\cite{Ad,GD}
The space $M_3$
 is the configuration space of $W_3$-gravity.
 We define in a canonical way the action of the second order differential
operators on $\Sl3$-opers. 
In this way we come to a Lie algebroid ${\cal A}_{M_3}$ over $\Sl3$-opers. 
The space of sections $\G({\cal A}_{M_3})$ is the
space of second order differential operators on $\Si_g$ without constant terms,
endowed by a Lie bracket. 
The algebroid ${\cal A}_{M_3}$ is lifted to the Hamiltonian algebroid  
${\cal A}^H_{{\cal R}_3}$ over
the phase space  ${\cal R}_3$ of $W_3$-gravity. 
The symplectic quotient of the phase space
is the so-called ${\cal W}_3$-geometry of $\Si_{g}$. Roughly speaking, this space is
a combination of the moduli of generalized complex structures
 and the spin 2 and 3 fields as the dual variables.
 Note that we deform the operator of complex structure $\bp$ by symmetric
combinations of vector fields $(\ve\p)^2$, in contrast with Ref.~\protect\citebk{BK}, where
the deformations of complex structures are carried out by the polyvector fields.
To define the $W_3$-geometry
we construct the BRST operator for the Hamiltonian algebroid. As it follows from
the general construction, it has the same structure as in the Lie algebra case.
It should be noted that the BRST operator for the $W_3$-algebras was constructed
in \cite{TM}. But here we construct the BRST operator for the different object
- the algebroid symmetries of $W_3$-gravity.
We explain our formulae and the origin of the algebroid
by the special gauge procedure of the $\Sl3$ Chern-Simons theory using an
approach developed in Ref.~\protect\citebk{BFK}.

The next example is the Poisson manifold ${\mathbb C}^4$ with 
Sklyanin algebra structure.\cite{Skl} As in the previous example,
we demonstrate that these algebroid structure can be derived from the
canonical brackets. The role of the Chern-Simons theory is played
by a free theory in $1+1$ dimension. It is a ''relativistic''
deformation of the Hitchin model for the elliptic Calogero-Moser
system. The Sklyanin algebra arises on an intermediate step of the symplectic
reduction.
Our construction is closed to the scheme proposed
in Ref.~\protect\citebk{AFM}. In conclusion, we perform the further symplectic reduction
with respect to the rest algebroid action
and obtain an integrable model with two degrees of freedom following Ref.~\protect\citebk{Do}.

\section{Lie algebroids}
\setcounter{equation}{0}

\subsection{Poisson manifolds} 

Let $M$ be a Poisson manifold endowed with
the Poisson bivector $\pi$. In local coordinates $x=(x_1,\ldots,x_n)$
 $\pi=\pi^{jk}$ and
$$
\pi^{jk}(x)=-\pi^{kj}(x), 
$$
\beq{2.1}
\p_i\pi^{jk}(x)\pi^{im}(x)+{\rm c.p.}(j,k,m)=0.
\eq

The Poisson brackets are defined on the space of smooth
functions ${\cal H}(M)$
\,\footnote{ ~The sums over repeated indices are understood in Sections 2-4.
 For simplicity we assume here that $M$ is a finite dimensional manifold,
 so later we shall consider the infinite dimensional case.}
$$
\{f(x),g(x)\}=\pi^{jk}\p_jf(x)\p_kg(x):=\lan\p f|\pi|\p g\ran,~~(\p_j=\p_{x^j})\,.
$$

The bivector allows to construct a map from  $f\in{\cal H}(M)$ 
 to the vector fields $V_f\in\G(TM)$ 
\beq{2.2}
f\to V_f= \pi^{jk}(x)\p_jf(x)\p_k=\lan\p f|\pi|\p\ran.
\eq
Let $\ve=(\ve_1,\ldots,\ve_n)$ be a section of $T^*M$.
Consider the vector field 
\beq{2.2a}
V_\ve=\lan\ve|\pi|\p\ran.
\eq
It acts on ${\cal H}(M)$ in the standard way
$$
\de_\ve f=i_{V_\ve}df=\lan\ve|\pi|df\ran.
$$
In particular,
\beq{dx}
\de_\ve x^k=\pi^{kj}(x)\ve_j.
\eq
One can define the brackets on the sections of $T^*M$ 
\beq{2.3}
\lb\ve,\ve'\rb=d\lan\ve|\pi(x)|\ve'\ran+\lan d\ve|\pi|\ve'\ran
+\lan\ve|\pi|d\ve'\ran,
\eq
or
$$
\lb\ve,\ve'\rb_m=\lan\ve|\p_m\pi|\ve'\ran+\de_\ve\ve'-\de_{\ve'}\ve.
$$
It follows from the Jacobi identity (\ref{2.1}), that the commutator
of the vector fields $V_\ve$ satisfies the identity
\beq{2.3a}
[ V_\ve,V_{\ve'}]=V_{\lb\ve,\ve'\rb}.
\eq

\subsection{Lie algebroids and Lie groupoids}

Now we can define the Lie algebroid over $M$. As we already mentioned
Lie algebroids is a generalization of bundles of Lie algebras over
a base $M$. We present
the definition of the Lie algebroid in general case.
Details of this theory can be find in 
 Ref.~\protect\citebk{Ma},  \protect\citebk{KM},  \protect\citebk{We}.
\begin{defi}
A {\em Lie algebroid} over a differential manifold $M$ is a vector bundle
${\cal A}\rar M$ with
a Lie algebra structure on the space of section $\G({\cal A})$
 defined by the brackets $\lb\ve_1,\ve_2\rb,~\ve_1,\ve_2\in\G({\cal A})$ and
a bundle map ({\em the anchor}) $\de :{\cal A}\to TM$, satisfying the following
conditions:\\
(i) For any $\ve_1,\ve_2\in\G({\cal A})$
\beq{2.4}
[\de_{\ve_1},\de_{\ve_2}]=\de_{\lb\ve_1\ve_2\rb},
\eq
(ii) For any $\ve_1,\ve_2\in\G({\cal A})$ and $f\in C^\infty(M)$
\beq{2.5}
\lb\ve_1,f\ve_2\rb=f\lb\ve_1,\ve_2\rb + (\de_{\ve_1} f)\ve_2.
\eq
\end{defi}
In other words, the anchor  defines a representation of the space of the
algebroid sections to the Lie algebra of vector fields
on the base $M$. The second condition is the Leibniz rule with respect
 to the multiplication of the sections by smooth functions.

A Lie algebra is a particular case of Lie algebroids if $M$ is a one-point space.
If the anchor map $\de_{\ve}\equiv 0$, then we have a bundle of Lie algebras
over $M$.

Let $\{e^j(x)\}$ be a basis of local sections $\G ({\cal A})$. Then  the brackets
are defined by the structure functions $f^{jk}_i(x)$ of the algebroid
\beq{5.1b}
\lb e^j,e^k\rb=f^{jk}_i(x)e^i,~~x\in V.
\eq
Using the Jacobi identity for the anchor action, we find
\beq{5.3}
C^n_{j,k,m}\de_{e_n}=0,
\eq
where
\beq{5.4}
C_n^{j,k,m}=(f^{jk}_i(x)f_n^{im}(x)+\de_{e^m}f^{jk}_n(x)+{\rm c.p.}(j,k,m))
\eq
Thus, (\ref{5.3}) implies {\em the anomalous Jacoby identity}
\beq{5.5}
f^{jk}_i(x)f_n^{im}(x)+\de_{e^m}f^{jk}_n(x)+{\rm c.p.}(j,k,m)=0
\eq
For the Lie algebras bundles the anomalous terms disappear.

In our case ${\cal A}=T^*M$ with the brackets (\ref{2.3}) defined on its
sections and the structure functions
$$
f^{jk}_i(x)=\p_i\pi^{jk}(x).
$$
 The anchor is also has a special form $\de_\ve=V_\ve$ (\ref{2.2a})
based on the Poisson bivector. 
The properties (\ref{2.4}) and (\ref{2.5}) follow from (\ref{2.3a}) 
and from (\ref{2.3}). We denote the algebroid $T^*M$ as ${\cal A}_M$.
This type of Lie algebroids was introduced in  
Ref.~\protect\citebk{Fu}, \protect\citebk{Ka}.

In generic case  Lie algebroids ${\cal A}$ can be integrated to  {\em  Lie groupoid}.\cite{Ba,Ma,We}
Again we present the general definition.
\begin{defi}
A Lie groupoid $G_M$ over a manifold $M$
is a pair of differential manifolds $(G,M)$,  two
differential mappings $l,r~: ~G_M\to M$ and a partially defined binary operation
(a product)
$(g,h)\mapsto g\cdot h $ satisfying the following conditions:\\
(i) It is defined only when $l(g)=r(h)$. \\
(ii) It is associative: $(g\cdot h)\cdot k=g\cdot (h\cdot k)$
whenever the products are defined.\\
(iii) For any $g\in G_M$ there exist the left and right identity elements $l_g$ and
$r_g$ such that $l_g \cdot g=g\cdot r_g=g$.\\
(iv) Each $g$ has an inverse $g^{-1}$ such that $g\cdot g^{-1}=l_g$ and
$g^{-1}\cdot g=r_g$.\\
\end{defi}
We denote an element of $g\in G_M$ by the triple $\ll x|g|y\gg$, 
where $x=l(g)$, $y=r(g)$.
Then  the product $g\cdot h $ is
$$
g\cdot h\rar \ll x|g\cdot h|z\gg=\ll x|g|y\gg\ll y|h|z\gg.
$$
An orbit of the groupoid in the base $M$ is defined as an equivalence
$x\sim y$ if $x=l(g),~y=r(g)$. There is the isotropy subgroup $G_x$ for $x\in M$.
$$
G_x=\{g\in G_M~|~l(g)=x=r(g)\}\sim\{\ll x|g|x\gg\}.
$$

\subsection{Central extension of Lie algebroids}

There is a straightforward generalization of this construction
to the infinite dimensional case. Consider 
 a disk $D=|z|\le 1$ and the space of anti-meromorphic maps\\
${\bf M} = \{X:D\to M\}$.
Define the Lie algebroid ${\cal A}_{\bf M}$ over the space 
${\bf M}$. Now the base of the algebroid is infinite-dimensional.
For simplicity we do not change the notion of the anchor action 
\beq{7.4}
\de_\ve X=\pi(X)|\ve\ran.
\eq
Here $\ve$ are the sections of $X^*(T^*M)$.

In the infinite dimensional case it is possible to
extend the anchor action (\ref{7.4})
\beq{7.4a}
\hat{\de}_\ve f(X)=\de_\ve f(X)+c(X,\ve),
\eq
by the one-cocycle 
\beq{7.7}
c(X,\ve)=\f1{2\pi}\oint\lan\ve|\bp X\ran.
\eq
Since 
\beq{7.7a}
{\de}_{\ve_1}c(X,\ve_2)-{\de}_{\ve_2}c(X,\ve_1)-c(X,\lb\ve_1,\ve_2\rb)=0,
\eq
(the cocycle property) one has
$$
[\hat{\de}_{\ve_1},\hat{\de}_{\ve_2}]=\hat{\de}_{\lb\ve_1,\ve_2\rb}.
$$
 We denote the Lie algebroid over ${\bf M}$ with the anchor (\ref{7.4a}) as
$\hat{\cal A}_{\bf M}$.
This construction is the generalization of the bundle of
the central extended loop Lie algebras with the loop parameter $\bz$.

 Consider the two-cocycle
$c(X;\ve_1,\ve_2)$ on $\G({\cal A}_{\bf M})$
\beq{5.9}
\de_{\ve_1}c(X;\ve_2,\ve_3)-
\de_{\ve_2}c(X;\ve_1,\ve_3)=0.
\eq
$$
+\de_{\ve_3}c(X;\ve_1,\ve_2)
-c(X;\lb \ve_1,\ve_2\rb,\ve_3)
+c(X;\lb \ve_1,\ve_3\rb,\ve_2)
-c(X;\lb \ve_2,\ve_3\rb,\ve_1).
$$
It allows to construct
the central extensions of brackets on $\G({\cal A})$
\beq{5.9a}
\lb(\ve_1,k_{1}),(\ve_2,k_{2})\rb_{c.e.}=
(\lb\ve_1,\ve_2\rb,c (x;\ve_1,\ve_2)).
\eq
The cocycle condition (\ref{5.9}) means that the new brackets
$\lb~,~\rb_{c.e.}$ satisfies AJI (\ref{5.5}).
The exact cocycles leads to the splitted extensions.

\section{Poisson sigma-model}

\subsection{Poisson sigma-model and Hamiltonian algebroids}

The manifold $M$ serves as the target space for the Poisson sigma model.
The space-time of the sigma model  is the disk $D$. 

Consider  the one-form on $D$ taking values in the pull-back by
$X$ of the cotangent bundle $T^*M$:
$$
\xi=(\xi_{1},\ldots,\xi_{n})
$$
Endow the space of fields ${\cal  R}=\{X,\xi\}$ with the canonical symplectic form
\beq{7.2a}
\om=\f1{2\pi}\oint\lan D\xi\we D X\ran.
\eq
Consider the set of the first-class constraints 
\beq{7.3}
F^j:=\bp X^j+(\pi(X)|\xi\ran)^j=0.
\eq
They generate the canonical transformations of $\om$
\beq{7.5}
\hat{\de}_\ve\xi_m=\frac{\de}{\de X^m}c(X,\ve)
+\lan\ve|\frac{\de}{\de X^m}\pi|\xi\ran=
\bp\ep_m+\lan\ve|\frac{\de}{\de X^m}\pi|\xi\ran,
\eq
and (\ref{7.4}). It means that we lift the anchor action on ${\cal R}$
 by means of the
cocycle (\ref{7.7}).
Equivalently, the
canonical transformations of a smooth functionals 
 can be described as 
$$
\de_{\ve_j} f(X,\xi)=\{h_{\ve_j},f(X,\xi)\}.
$$
Here the Poisson brackets are inverse to the symplectic form $\om$ (\ref{7.2a})
and
\beq{7.8}
h_{\ve_j}=\f1{2\pi}\int\ve_jF^j,
\eq
(no summation on $j$). 
Again, due to (\ref{7.7a})
\beq{7.9}
\{h_\ve,h_{\ve '}\}=h_{\lb\ve,\ve '\rb}.
\eq

Summarizing, we have defined the symplectic manifold ${\cal R}\{\xi,X\}$ 
and the bundle
${\cal A}^H_{\cal R}$ over ${\cal R}$ with the sections 
$\ve\in\G({\cal A}^H_{\cal R})$. 

For a general symplectic manifold ${\cal R}$ this construction leads 
to the Hamiltonian algebroid
\cite{LO}
\begin{defi}
${\cal A}^H_{\cal R}$ is a {\em Hamiltonian  algebroid} over a symplectic
manifold ${\cal R}$ if there is
a bundle map from ${\cal A}^H_{\cal R}$ to the Lie algebra
on $C^\infty({\cal R})$: $\ve\to h_\ve$, (i.e. $f\ve\to fh_\ve$ for
$f\in C^\infty({\cal R})$)
satisfying the following conditions:\\
(i) For any $\ve_1,\ve_2\in\G ({\cal A}^H_{\cal R})$ and $x\in {\cal R}$
\beq{5.12}
\{ h_{\ve_1},h_{\ve_2}\}=h_{\lb\ve_1,\ve_2 \rb}.
\eq\\
(ii) For any $\ve_1,\ve_2\in\G ({\cal A}^H_{\cal R})$ and $f\in C^\infty({\cal R})$
\beq{5.13}
\lb\ve_1,f\ve_2\rb=f\lb\ve_1,\ve_2\rb+\{h_{\ve_1},f\}\ve_2.
\eq
\end {defi}
The both conditions are similar to the defining properties of the
Lie algebroids (\ref{2.4}),(\ref{2.5}). For general Hamiltonian algebroids
the Jacobi identity takes the form
\beq{5.14a}
f^{jk}_i(x)f_n^{im}(x)+\{h_{e^m},f^{jk}_n(x)\}+
E^{j,k,m}_{[ln]}h_{\ve^l}+c.p.(j,k,m)=0,
\eq
where $E^{j,k,m}_{[ln]}$ is an antisymmetric tensor in $[ln]$.
This structure arises in the Hamiltonian systems with the first class constraints
 and leads to the open algebra of arbitrary rank.\cite{HT}

We have constructed a special Hamiltonian algebroid 
${\cal A}^H_{\cal R}$ with the brackets
(\ref{2.3}), the base ${\cal R}$ equipped with the canonical form (\ref{7.2a})
and the anchor (\ref{7.4}), (\ref{7.5}). 
In this case the last term in (\ref{5.14a}) 
is absent
 and we come to the anomalous Jacobi identity similar to (\ref{5.5})
\beq{5.14}
f^{jk}_i(x)f_n^{im}(x)+\{h_{m},f^{jk}_n(x)\}=0,
\eq
where $h_m$ is  defined by (\ref{7.8}). This identity plays the crucial role 
in the BRST construction.

\subsection{BRST construction for Hamiltonian algebroids}

Here we consider the BRST construction for general Hamiltonian algebroids.
 Let ${\cal A}^{H*}$ be the dual bundle.
Its  sections $\eta \in\G({\cal A}^{H*}_{\cal R})$ are the odd fields called
 {\em the ghosts}.
Let 
$$
h_{e^j}=\f1{2\pi}\oint\lan\eta_j|F(x)\ran,
$$
 where $\{\eta_j\}$ is a basis in
$\G({\cal A}^{H*})$ and $F(x)=0$ are the moment constraints, generating the
canonical algebroid action on ${\cal R}$.
 Introduce another type of odd variables ({\em the ghost momenta})
${\cal P}^j\in \G({\cal A}^{H}_{\cal R}),~j=1,2,\ldots$
 dual to the ghosts $\eta_k,~k=1,2,\ldots$.
 We attribute the ghost
 number one to the ghost fields gh$(\eta)=1$, minus one to the ghost momenta
gh$({\cal P})=-1$ and gh$(x)=0$ for $x\in {\cal R}$.
Introducing
 In addition to the non-degenerate Poisson structure on ${\cal R}$
 we introduce the Poisson brackets
\beq{5.19}
\{\eta_j,{\cal P}^k\}=\de_j^k,~~\{\eta^j,x\}=\{{\cal P}_k,x\}=0.
\eq

Thus all fields are incorporated in the graded Poisson superalgebra
$$
{\cal BFV}=\left(
\G(\wedge^\bullet  ({\cal A}^{H*}_{\cal R}\oplus{\cal A}^{H}_{\cal R})
\right)\otimes C^\infty({\cal R}).
=\G(\wedge^\bullet {\cal A}^{H*}_{\cal R})
\otimes\G(\wedge^\bullet {\cal A}^{H}_{\cal R})\otimes C^\infty({\cal R}).
$$
({\em the Batalin-Fradkin-Vilkovitsky (BFV) algebra}).

There exists a nilpotent operator on $Q$ on ${\cal BFV}$ 
$(Q^2=0,~gh(Q)=1)$
({\em the BRST operator}) transforming ${\cal BFV}$ into the BRST complex.
The cohomology of ${\cal BFV}$ complex give rise to the structure of
the classical reduced phase space ${\cal R}^{red}$. Namely $H^0(Q)$
is identified with the classical observables.

Represent the action of $Q$ as the Poisson brackets:
$$
Q\psi=\{\psi,\Om\},~~\psi,\Om\in {\cal BFV}.
$$
Due to the Jacobi identity for the Poisson brackets the nilpotency of $Q$
is equivalent to
$$
\{\Om,\Om\}=0.
$$
Since $\Om$ is odd, the brackets are symmetric.
 For a generic Hamiltonian systems $\Om$ has the form of infinite series
in the ghost ${\cal P}^k$ expansion
$$
\Om=h_\eta+\f1{4\pi}\oint\lan\lb\eta,\eta'\rb|{\cal P}\ran+...,~
(h_\eta=\oint\lan\eta|F\ran).
$$
The order of ${\cal P}$
in $\Om$ is called {\em the rank} of the BRST operator $Q$.
For generic Hamiltonian algebroids with the anomalous Jacobi identity
(\ref{5.14a}) $Q$ can have the infinite rank.

If ${\cal A}$ is a Lie algebra bundle defined along with its canonical
action on ${\cal R}$ then $Q$ has the rank one or less. In this case
the BRST operator $Q$ is the extension of the Cartan-Eilenberg operator
giving rise to the cohomology of ${\cal A}$ with coefficients in
$C^\infty({\cal R})$ and the first two terms in the previous expression 
provide the nilpotency of $Q$. 

The distinguish property of ${\cal A}^H_{\cal R}$
is that  $\Om$ has the rank one, though the Jacobi identity has additional
terms (\ref{5.14}) in compare with the Lie algebras \cite{LO}
\beq{5.20}
\Om=
\f1{2\pi}\oint\lan\eta|F\ran+\f1{4\pi}\oint\lan\lb\eta,\eta'\rb|{\cal P}\ran
\eq

\section{$W_3$-gravity}
\setcounter{equation}{0}

\subsection{Lie algebroid over $\Sl3$-opers}

In our approach the starting point to the $W_3$-gravity
is the space infinite dimensional Poisson space of the third order
differential operators with smooth coefficients
on a Riemann curve $\Si_n$ (the $\Sl3$-opers). The curve $\Si_n$
play the role of the
space part of the $W_3$-gravity, while the time is just $\mathbb R$.
Locally $\Sl3$-opers are defined as 
\beq{6.2}
\p^3-T\p-W:~\Om^{(-1,0)}(\Si_{g})\to\Om^{(2,0)}(\Si_{g}),~~(\p=\p_z),
\eq
where $\Om^{(a,b)}(\Si_{g})$ is the space of section of smooth $(a,b)$ forms
on $\Si_g$.
In fact, they can be defined globally starting from $\Sl3$ bundles
over $\Si_g$. 
The connection on the  $\Sl3$ bundle has the special form corresponding to
(\ref{6.2})
\beq{6.1}
\nabla=\p -\thmat{0}{1}{0}{0}{0}{1}{W}{T}{0}.
\eq
The  $\SLN$ opers were introduced in  Ref.~\protect\citebk{Te}. The general case was considered  in  Ref.~\protect\citebk{BD}.

The space of $\Sl3$-opers plays the role of the Poisson manifold
\mbox{${\bf M}=M_3=\{W,T\}$,} where the brackets are defined as
\beq{6.9a}
\{T(z),T(w)\}
=\left(-2\ka\p^3+2T(z)\p+\p T(z)\right)\de(z-w),\eq
\beq{6.10a}
\{T(z),W(w)\}=\left(-\ka\p^4+T(z)\p^2+3W(z)\p+\p W(z)\right)\de(z-w),
\eq
\beq{6.12a}
\{W(z),W(w)\}=
\eq
$$
+\left(\frac{2}{3}k\p^5-\frac{4}{3}T(z)\p^3-
2\p T(z)\p^2+\left(\frac{2}{3}(2-\ka)T(z)^2-2\p^2T(z)+2\p W(z)\right)\p
\right.
$$
$$
\left. +\left(\p^2W(z)-\frac{2}{3}\p^3T(z)+\frac{2}{3}(2-\ka)T(z)\p T(z)\right)
\right)\de(z-w).
$$
These brackets are defined in terms of coordinates 
in a neighborhood of a non-contractable contour on $\Si_g$.
The parameter $\ka$ comes from the central extension 
 of the Lie brackets on the
sections of the algebroid ${\cal A}_3$ and we have $2g$ non-equivalent
extensions (see below (4.20) and Remark 4.1). For $\ka\neq 2$
 the Poisson brackets $\{W(z),W(w)\}$ have two
quadratic terms $T(z)^2$ and $\frac{2}{3}T(z)\p T(z))$.
In what follows we take $\ka=1$.
 Note, that $\bz$ dependence is hidden these relations. 
In this way the
Poisson structure on $M_3$
 is just the global version of Adler-Gelfand-Dikii brackets.\cite{Ad,GD} 

Define the Lie algebroid ${\cal A}_3$ over $M_3$.
Its sections  ${\cal D}_2 =\G({\cal A}_3)$ are the 
 second order differential operators on
$\Si_{g}$ without constant terms. On a disk ${\cal D}_2$ can be trivialized
and the sections are represented as
$$
\ve^{(1)}=\ve^{(1)}(z,\bz)\frac{\p}{\p z},~~
\ve^{(2)}=\ve^{(2)}(z,\bz)\frac{\p^2}{\p z^2},
$$
$$
\ve^{(1)}\in {\cal D}^1,~~\ve^{(2)}\in {\cal D}^2,~~
{\cal D}_2={\cal D}^1\oplus{\cal D}^2.
$$
The second order differential operators do not generate
a closed algebra with respect to the standard commutators.
Moreover, they cannot be defined
invariantly on Riemann curves in contrast with the first order
differential operators.
We introduce a new brackets that goes around the both disadvantages.
The antisymmetric brackets on ${\cal D}_2$ are defined 
by means the Poisson brackets (\ref{6.9a}),(\ref{6.10a}),(\ref{6.12a}) 
according to the general prescription (\ref{2.3}).
\beq{6.6}
\lb\ve^{(1)}_1,\ve^{(1)}_2\rb=\ve^{(1)}_1\p\ve^{(1)}_2-\ve^{(1)}_2\p\ve^{(1)}_1.
\eq
\beq{6.7}
\lb\ve^{(1)},\ve^{(2)}\rb=\left\{
\begin{array}{cl}
-\ve^{(2)}\p^2\ve^{(1)}, &\in{\cal D}^1\\
-2\ve^{(2)}\p\ve^{(1)}+\ve^{(1)}\p\ve^{(2)},  &\in{\cal D}^2
\end{array}
\right.
\eq
\beq{6.8}
\lb\ve^{(2)}_1,\ve^{(2)}_2\rb=\left\{
\begin{array}{cl}
\frac{2}{3}[\p(\p^2-T)\ve^{(2)}_1]\ve^{(2)}_2 -
 \frac{2}{3}[\p(\p^2-T)\ve^{(2)}_2]\ve^{(2)}_1   , &\in{\cal D}^1\\
\ve^{(2)}_2\p^2\ve^{(2)}_1-\ve^{(2)}_1\p^2\ve^{(2)}_2,  &\in{\cal D}^2
\end{array}
\right.
\eq
The brackets (\ref{6.6}) are the standard Lie brackets of vector fields 
and therefore ${\cal D}^1$ is the Lie subalgebra of  ${\cal D}_2$.
The structure functions in (\ref{6.8}) depend on  the projective connection $T$.

Now consider the bundle map ${\cal A}_3$ to $TM_3$ defined by the
anchor \\
$\de_{\ve^{(j)}}=\lan\ve^{(j)}|\pi|  \frac{\de}{\de T}\ran$,
or $\de_{\ve^{(j)}}=\lan\ve^{(j)}|\pi|  \frac{\de}{\de W}\ran$
\beq{6.9}
\de_{\ve^{(1)}}T=-2\p^3\ve^{(1)}+2T\p\ve^{(1)}+\p T\ve^{(1)},
\eq
\beq{6.10}
\de_{\ve^{(1)}}W=-\p^4\ve^{(1)}+3W\p\ve^{(1)}+\p W\ve^{(1)}+T\p^2\ve^{(1)},
\eq
\beq{6.11}
\de_{\ve^{(2)}}T=\p^4\ve^{(2)}-T\p^2\ve^{(2)}+(3W-2\p T)\p\ve^{(2)}+
(2\p W-\p^2T)\ve^{(2)},
\eq
\beq{6.12}
\de_{\ve^{(2)}}W=\frac{2}{3}\p^5\ve^{(2)}-\frac{4}{3}T\p^3\ve^{(2)}-
2\p T\p^2\ve^{(2)}
\eq
$$
+(\frac{2}{3}T^2-2\p^2T+2\p W)\p\ve^{(2)}+
(\p^2W-\frac{2}{3}\p^3T+\frac{2}{3}T\p T)\ve^{(2)}.
$$

The algebroid structure follows from the identity
$$
[\de_{\ve^{(j)}_1},\de_{\ve^{(k)}_2}]=\de_{\lb\ve^{(j)}_1,\ve^{(k)}_2 \rb},~~
(j,k=1,2).
$$

The Jacobi identity (\ref{5.5}) in ${\cal A}_3$ takes the form
\beq{6.14}
\lb\lb\ve^{(2)}_1,\ve^{(2)}_2\rb,\ve^{(2)}_3\rb^{(1)}-
(\ve^{(2)}_1\p\ve^{(2)}_2-\ve^{(2)}_2\p\ve^{(2)}_1)\de_{\ve^{(2)}_3}T+
{\rm c.p.}(1,2,3)
=0,
\eq
\beq{6.15}
\lb\lb\ve^{(2)}_1,\ve^{(2)}_2\rb,\ve^{(1)}_3\rb^{(1)}+{\rm c.p.}(1,2,3)=
(\ve^{(2)}_1\p\ve^{(2)}_2-\ve^{(2)}_2\p\ve^{(2)}_1)\de_{\ve^{(1)}_3}T.
\eq
The brackets here correspond to the product of
structure functions in the left hand side
of (\ref{5.5}) and the superscript $(1)$ corresponds to the ${\cal D}^1$ component.
For the  rest  brackets the Jacobi identity is the standard one.

The origin of the brackets and the anchor representations follow from the
matrix description of $\Sl3$-opers (\ref{6.1}).
Let $E_3$ be the principle $\Sl3$-bundle over $\Si_g$.
Consider the set ${\cal G}_3$ of automorphisms of the bundle $E_3$
\beq{6.15a}
A\to f^{-1}\p f-f^{-1}Af
\eq
that preserve the $\Sl3$-oper structure
\beq{6.16}
f^{-1}\p f-
f^{-1}A(W,T)f=A(W',T'),
\eq
$$
 A(W,T)=\thmat{0}{1}{0}{0}{0}{1}{W}{T}{0}.
$$
It is clear that ${\cal G}_3$ is the Lie groupoid over  $M_3=\{W,T\}$ with
$l(f)=(W,T)$, $~r(f)=(W',T')$, $~f\to \ll W,T|f|W',T'\gg $.
The left identity map is the $\Sl3$  subgroup of ${\cal G}_3$
$$
P\exp\left
(-\int^z_{z_0} A(W,T)\right)
\cdot C\cdot P\exp\left(\int^z_{z_0} A(W,T)
\right),
$$
where $C$ is an arbitrary matrix from $\Sl3$.
The right identity map has the same
form with $(W,T)$ replaced by $(W',T')$.

The local version of (\ref{6.16}) takes the form
\beq{6.17}
\p X-\left[\thmat{0}{1}{0}{0}{0}{1}{W}{T}{0},X\right]=
\thmat{0}{0}{0}{0}{0}{0}{\de W}{\de T}{0}.
\eq
It is the sixth order linear differential system for the matrix elements of the
traceless matrix $X$. The matrix elements $x_{j,k}\in\Om^{(j-k,0)}(\Si_{g})$
 depend on two arbitrary fields $x_{23}=\ve^{(1)},~x_{13}=\ve^{(2)}$. The
solution takes the form
\beq{6.17b}
X(\ve^{(1)},\ve^{(2)})=\thmat{x_{11}}{x_{12}}{\ve^{(2)}}{x_{21}}{x_{22}}
{\ve^{(1)}}{x_{31}}{x_{32}}{x_{33}},
\eq
$$
x_{11}=\frac{2}{3}(\p^2-T)\ve^{(2)}-\p\ve^{(1)},~
x_{12}=\ve^{(1)}-\p\ve^{(2)},
$$
$$
x_{21}=\frac{2}{3}\p(\p^2-T)\ve^{(2)}-\p^2\ve^{(1)}+W\ve^{(2)},~
x_{22}=-\frac{1}{3}(\p^2-T)\ve^{(2)},
$$
$$
x_{31}=\frac{2}{3}\p^2(\p^2-T)\ve^{(2)}-\p^3\ve^{(1)}+\p(W\ve^{(2)})+W\ve^{(1)},
$$
$$
x_{32}=\frac{1}{3}\p(\p^2-T)\ve^{(2)}-\p^2\ve^{(1)}+W\ve^{(2)}+T\ve^{(1)},
$$
$$
x_{33}=-\frac{1}{3}(\p^2-T)\ve^{(2)}+\p\ve^{(1)}.
$$
The matrix elements of the commutator $[X_1,X_2]_{13}$, $[X_1,X_2]_{23}$
give rise to the brackets (\ref{6.6}), (\ref{6.7}), (\ref{6.8}). Simultaneously,
from (\ref{6.17}) one obtain the anchor action (\ref{6.9})-(\ref{6.12}).

There is a nontrivial cocycle corresponding to $H^1({\cal A}_3)$
with two components
\beq{6.17a}
c_\al(\ve^{(j)}_1,\ve^{(k)}_2)=\oint_{\ga_\al}\la(\ve^{(j)}_1,\ve^{(k)}_2),
~~(j,k=1,2),
\eq

The cocycle allows to shift the anchor action
$$
\hat{\de}_{\ve^{(j)}}f(W,T)=
\int_{\Si_{g}}\lan\de_{\ve^{(j)}}W|\frac{\de f}{\de W}\ran+
\int_{\Si_{g}}\lan\de_{\ve^{(j)}}T|\frac{\de f}{\de T}\ran+c^{(j)},~~(j=1,2).
$$

There exists the $2g$ central extensions $c_\al$ of the algebra ${\cal D}_2$,
provided by the nontrivial cocycles from  $H^2({\cal A}_3,M_3)$. They are the
non-contractible contour integrals $\ga_\al$
\beq{6.16a}
c_\al(\ve^{(j)}_1,\ve^{(k)}_2)=\oint_{\ga_\al}
\la(\ve^{(j)}_1,\ve^{(k)}_2),
~~(j,k=1,2),
\eq
where
$$
\la(\ve^{(1)}_1,\ve^{(1)}_2)=\ve_1^{(1)}\p^3\ve_2^{('1)},~~
\la(\ve^{(1)}_1,\ve^{(2)}_2)=\ve^{(1)}_1\p^4\ve^{(2)}_2,
$$
$$
\la(\ve^{(2)}_1,\ve^{(2)}_2)=
$$
$$
\frac{2}{3}\left(
-(\p^2-T)\ve_1^{(2)}\p(\p^2-T)\ve_2^{(2)}
+\p^2\ve_1^{(2)}\p(\p^2-T)\ve_2^{(2)}+\p^2\ve_2^{(2)}\p(\p^2-T)\ve_1^{(2)}
\right).
$$
It can be proved that $sc^j=0$
(\ref{5.9}) and that $c^j$ is not exact. 
These cocycles allow to construct the central extensions of $\hat{\cal A}_3$:
$$
\lb(\ve_1^{(j)},\sum_\al k^{(j)}_\al),(\ve_2^{(m)},\sum_\al k^{(m)}_\al)\rb_{c.e.}=
(\lb\ve_1^{(j)},\ve_2^{(m)}\rb,\sum_\al c_\al(\ve_1^{(j)},\ve_2^{(m)})).
$$
\begin{rem}
As in the case of the Lie algebras
 the Poisson structure (\ref{6.9a}),
(\ref{6.10a}) and (\ref{6.12a}) can be read off from the Lie brackets
(\ref{6.6}), (\ref{6.7}), (\ref{6.8}) along with their central extensions
(\ref{6.16a})
according with the Lie brackets for $H_1,H_2\in C^{\infty}(M_3)$
$$
\{H_1,H_2\}=\oint_{\ga_\al}\lan\lb\de H_1,\de H_2\rb X\ran+
k\oint_{\ga_\al}\la(\de H_1,\de H_2),
$$
where $X=W,T$ and $\de H=\de H/\de X\in {\cal D}_2$.
\end{rem}

\subsection{Phase space of $W_3$-gravity and BRST operator}

The dual fields to $T$ and $W$ are the Beltrami differentials
$\mu\in\Om^{(-1,1)}(\Si_{g})$ and the differentials 
$\rho\in\Om^{(-2,1)}(\Si_{g})$. Along with $T$ and $W$ they generate
the space   ${\cal R}_3$. It is the classical phase space for the 
$W_3$-gravity.\cite{P,BFK,GLM} The symplectic form on ${\cal R}_3$ 
has the canonical form
\beq{6.20}
\om=\int_{\Si_{g}}D T\we D\mu+D W\we D\rho.
\eq

According with the general theory the anchor (\ref{6.9})-(\ref{6.12})
can be lifted from $M_3$ to ${\cal R}_3$.
This lift is nontrivial owing to the cocycle (\ref{6.17a}).
It follows from (\ref{7.5}) that the anchor action on $\mu$ and
$\rho$ takes the form
\beq{6.21}
\de_{\ve^{(1)}}\mu=-\bar{\p}\ve^{(1)}-\mu\p\ve^{(1)}+\p\mu\ve^{(1)}-
\rho\p^2\ve^{(1)},
\eq
\beq{6.22}
\de_{\ve^{(1)}}\rho=-2\rho\p\ve^{(1)}+\p\rho\ve^{(1)},
\eq
\beq{6.23}
\de_{\ve^{(2)}}\mu=\p^2\mu\ve^{(2)}-\frac{2}{3}\left[(\p(\p^2-T)\rho)\ve^{(2)}
-(\p(\p^2-T)\ve^{(2)})\rho\right],
\eq
\beq{6.24}
\de_{\ve^{(2)}}\rho=-\bar{\p}\ve^{(2)}+(\rho\p^2\ve^{(2)}-\p^2\rho\ve^{(2)})
+2\p\mu\ve^{(2)}-\mu\p\ve^{(2)}.
\eq
In this way we have defined the Hamiltonian Lie algebroid over ${\cal R}_3$ with 
the section  ${\cal D}_2$, the brackets (\ref{6.6}),(\ref{6.7}),(\ref{6.8}),
and the anchor (\ref{6.9})-(\ref{6.12}), (\ref{6.21})-(\ref{6.24}).

There are two Hamiltonians, defining by the anchor and by the cocycle 
$$
h^{(1)}=\int_{\Si_{g}}(\mu\de_{\ve^{(1)}}T+\rho\de_{\ve^{(1)}}W)+c^{(1)},~~
h^{(2)}=\int_{\Si_{g}}(\mu\de_{\ve^{(2)}}T+\rho\de_{\ve^{(2)}}W)+c^{(2)}.
$$
After the integration by part they take the form
$$
h^{(1)}=\int_{\Si_{g,n}}\ve^{(1)}F^{(1)},
$$
$$
h^{(2)}=\int_{\Si_{g,n}}\ve^{(2)}F^{(2)},
$$
where $F^{(1)}\in\Om^{(2,1)}(\Si_{g,n})$, $F^2\in\Om^{(3,1)}(\Si_{g,n})$
\beq{F1}
F^{(1)}=-\bp T-\p^4\rho+T\p^2\rho-(3W-2\p T)\p\rho-
\eq
$$
-(2\p W-\p^2 T)\rho+2\p^3\mu-2T\p\mu-\p T\mu,
$$
\beq{F2}
F^{(2)}=-\bp W-\frac{2}{5}\p^5\rho+\frac{4}{3}T\p^3\rho+2\p T\p^2\rho
(-\frac{2}{3}T^2+2\p^2 T-2\p W)\p\rho+
\eq
$$
+(-\p^2W+\frac{2}{3}\p^3T-\frac{2}{3}T\p T)\rho+\p^4\mu
-3W\p\mu-\p W\mu-T\p^2\mu.
$$
They carry out the moment map
$$
m=(m^{(1)}=F^{(1)},m^{(2)}=F^{(2)}): {\cal R}_3\to \G^*({\cal A}_3).
$$
We assume that $m^{(1)}=0,~m^{(2)}=0$.
Then the coadjoint action of ${\cal D}_2$ preserve $m=(0,0)$.
The moduli space  ${\cal W}_3$ of the $W_3$-gravity
( $W_3$-geometry) is the symplectic quotient with respect to the groupoid
${\cal G}_3$ action
$$
{\cal W}_3={\cal R}_3//{\cal G}_3=\{F^1=0,F^2=0\}/{\cal G}_3.
$$
The space ${\cal W}_3$ is finite-dimensional.
It has dimension
$$
\dim{\cal W}_3=16(g-1).
$$

The prequantization of ${\cal W}_3$ can be realized in the space of sections
of a linear bundle ${\cal L}$ over the space of orbits
 $\ti{M}_3\sim M_3/{\cal G}_3$. 
The sections are functionals $\Psi(T,W)$ on $M_3$ satisfying
the following conditions
$$
\hat{\delta}_{\ve^{(j)}}\Psi(T,W):=
\lan\de_{\ve^{(j)}}W|\frac{\de \Psi}{\de W}\ran+
\lan\de_{\ve^{(j)}}T|\frac{\de \Psi}{\de T}\ran+c^{(j)}\Psi=0,~~(j=1,2).
$$
Presumably, the bundle ${\cal L}$ can be identified with the determinant bundle
 $\det(\p^3-T\p-W)$.

The moment equations $F^{(1)}=0,~F^{(2)}=0$ are the consistency conditions
for the linear system
\beq{6.26a}
\left\{
\begin{array}{l}
(\p^3-T\p-W)\psi(z,\bz)=0,\\
\left(\bp +(\mu-\p\rho)\p +\rho\p^2
+\frac{2}{3}(\p^2-T)\rho-\p\mu
\right)\psi(z,\bz)=0,
\end{array}
\right.
\eq
where $\psi(z,\bz)\in\Om^{(-1,0)}(\Si_{g})$. 
The last equation  represents the deformation of the antiholomorphic
operator $\bp$ by the first order operator $\mu\p$ and
by the second order differential
operator $\rho\p^2$. The left hand side is the exact form of the deformed operator
when it acts on $\Om^{(-1,0)}(\Si_{g})$. This deformation cannot
be supported by the structure of a Lie algebra  and one leaves with
the Hamiltonian algebroid symmetries.

Instead of the symplectic reduction one can apply the BRST construction.
Then cohomology of the moduli space  ${\cal W}_3$ are
isomorphic to
$H^j(Q)$. To construct the BRST complex we introduce the ghosts
fields $\eta^{(1)},\eta^{(2)}$ and their momenta ${\cal P}^{(1)},{\cal P}^{(2)}$.
Then it follows from the general construction that for
$$
\Om=\sum_{j=1,2}h^{(j)}(\eta^{(j)})+
\oh\sum_{j,k,l=1,2}\int_{\Si_{g,n}}(\lb\eta^{(j)},\eta^{(k)}\rb{\cal P}^{(l)}).
$$
The operator  $QF=\{F,\Om\}$ is nilpotent and defines the BRST cohomology
in the complex
$$
\bigwedge{}^\bullet({\cal D}_2\oplus{\cal D}_2^*)\otimes C^\infty({\cal R}_3).
$$

\subsection{Chern-Simons description}

Here we follow the approach proposed in  Ref.~\protect\citebk{BFK}.
Consider the Chern-Simons functional on $\Si_{g}\oplus {\bf R}^+$
$$
S=\int_{\Si_{g}\oplus {\bf R}}\tr
\left({\cal A}d{\cal A}+\frac{2}{3}{\cal A}^3\right),~~
{\cal A}=(A,\bA,A_t).
$$

In the hamiltonian picture the components
$A,\bA$ are elements of the phase space ${\cal R}_{\Sl3}$
 with the symplectic form
$$
\int_{\Si_{g}}DA\wedge D\bA,
$$
 while $A_t$ is the Lagrange multiplier for the first class constraints
\beq{8.5a}
F(A,\bA):=\bp A-\p \bA +[\bA,A]=0.
\eq

The phase space ${\cal R}_3$ can be derived from the phase space
 of the Chern-Simons theory ${\cal R}_{\Sl3}$. 
The flatness condition (\ref{8.5a}) generates
the gauge transformations
\beq{5.31}
A\to f^{-1}\p f+f^{-1}Af,~~\bA\to f^{-1}\bp f+f^{-1}\bA f,
~~f\in G_{\Sl3}
\eq
The result of the gauge fixing with respect to the whole gauge group $G_{\Sl3}$
is the moduli
space ${\cal M}^{flat}_3$ of the flat $\Sl3$ bundles over $\Si_{g}$.

Let $P$ be the maximal parabolic subgroup of $\Sl3$ of the form
$$
P=\thmat{*}{*}{0}{*}{*}{0}{*}{*}{*},
$$
and $G_P$ be the corresponding gauge group.
We partly fix the gauge with respect to $G_P$.
A generic connection $\nabla$ can be gauge transformed by $f\in G_P$ to
  the form (\ref{6.1}).

The form of $\bA$ can be read off from (\ref{8.5a})
\beq{5.32}
\bA=\thmat{a_{11}}
{a_{12}}
{-\rho}
{a_{21}}
{a_{22}}
{-\mu}
{a_{31}}
{a_{32}}
{a_{33}}
\eq
$$
a_{11}=-\frac{2}{3}(\p^2-T)\rho+\p\mu,~~
a_{12}=-\mu+\p\rho,
$$
$$
a_{21}=-\frac{2}{3}\p(\p^2-T)\rho+\p^2\mu-W\rho,~~
a_{22}=\frac{1}{3}(\p^2-T)\rho,
$$
$$
a_{31}=-\frac{2}{3}\p^2(\p^2-T)\rho+\p^3\mu-\p(W\rho)-W\mu,
$$
$$
a_{32}=-\frac{1}{3}\p(\p^2-T)\rho+\p^2\mu-W\rho-T\mu,~~
a_{33}=\frac{1}{3}(\p^2-T)\rho-\p\mu.
$$
The flatness (\ref{8.5a})  for the special choice $A$ (\ref{6.1}) and
$\bA$ (\ref{5.32}) gives rise to
the moment constraints $F^{(2)}=0,~F^{(1)}=0$ (\ref{F1}),(\ref{F2}).
 Namely, one has
$F(A,\bA)|_{(3,1)}=F^{(2)}$ (\ref{F1}), $F(A,\bA)|_{(2,1)}=F^{(1)}$ (\ref{F2}),
while the other matrix
elements of $F(A,\bA)$ vanish identically.
In this way, we come to the matrix description of the moduli space
${\cal W}_3$.

The groupoid action on $A,\bA$ plays the role of the rest gauge
transformations that complete the $G_P$ action to the $G_{\Sl3}$ action.
The algebroid symmetry arises in this theory as a result of the partial
gauge fixing by $G_{P}$. Thus we come to the following diagram

\bigskip
$$
\begin{array}{rcccl}
           &\fbox{${\cal R}_{\Sl3}$}&                       &                    &\\
           &       |                      &\searrow{G_{P}}&           &\\
G_{\Sl3} &        |                     &                    &\fbox{${\cal R}_3$}&\\
           &\downarrow    &                   &\downarrow &\G({\cal G}_3)\\
           &\fbox{${\cal M}^{flat}_{\Sl3}$}&       &\fbox{${\cal W}_3$}&\\
\end{array}
$$
\bigskip

The tangent space to ${\cal M}^{flat}_{\Sl3}$ at the point $(A=0,\bA=0)$
coincides with the tangent space to ${\cal W}_3$ at the point
$(W=0,T=0,\mu=0,\rho=0)$. Their dimension is $16(g-1)$. But their global
structure is different and the diagram cannot be closed by the horizontal
isomorphisms. The interrelations between ${\cal M}^{flat}_{\Sl3}$ and ${\cal W}_3$
were analyzed by Hitchin.\cite{H2}

\section{Sklyanin algebra}
\setcounter{equation}{0}

\subsection{Hamiltonian algebroid for Sklyanin algebra}

We start from the Poisson bivector defined on a ${\mathbb C}^4$ with
generators\\ $(S_0,\vec{S}=(S_\al),~\al=1,2,3)$
\footnote{~In this section to reconcile with the accepted notations
of the Sklyanin algebra we do not comply with the positions of the upper
and lower indices.}  
\beq{0}
\{S_\al,S_0\}=2J_{\be\ga}S_\be S_\ga,~~\{S_\al,S_\be\}=-2\ep_{\al\be\ga}S_0S_\ga,
\eq
where $J_{\be\ga}=\wp_\be-\wp_\ga$,   $\wp_\al=e_\al$,
\footnote{~We use the standard notations
of the elliptic functions.\cite{Akh}}
 and
\beq{0a}
e_1=\wp\left(\oh\right),~e_2=\wp\left(\frac{1+\tau}{2}\right),~
e_3=\wp\left(\frac{\tau}{2}\right),
\eq
and $\wp=\wp(u;\tau)$ is the Weierstrass function. 
In this way
the algebra is parameterized by the modular parameter $\tau$.
The Jacobi identity is provided by the identity
$e_1+e_2+e_3=0$. 
It is the Sklyanin algebra ${\cal S}$.\, \cite{Skl}
It has two Casimirs
\beq{K}
K_0=\sum S_\al^2~~,K_1=S_0^2+\sum\wp_\al S_\al^2.
\eq
The Lie algebroid brackets (\ref{2.3}) for 
$(\ve_0,\vec {\ve})\in T^*{\cal S}$ take the form
$$
 \lb\ve_\be,\ve_0\rb_\al=2J_{\be\al}S_\be\ve_\al,
$$
\beq{0b}
 \lb\ve_\al,\ve_\be\rb_0=-2\ep_{\al\be\ga}S_\ga\ve_\al\ve_\be,
\eq
$$
 \lb\ve_\al,\ve_\be\rb_\ga=-2\ep_{\al\be\ga}S_0\ve_\al\ve_\be.
$$

It is convenient to arrange the Poisson brackets on ${\cal S}$ in the 
form of the classical Yang-Baxter equation.
Introduce three functions $\vf_\al(z)$ depending on
a formal (spectral) parameter $z$ living on the elliptic
curve $\Si_\tau={\mathbb C}/({\mathbb Z}+\tau{\mathbb Z})$
\beq{0c}
\vf_1(z)=\phi\left(\oh,z\right),~
\vf_2(z)=-e^{\pi i\tau}\phi\left(\frac{1+\tau}{2},z\right),~
\vf_3(z)=e^{\pi i\tau}\phi\left(\frac{\tau}{2},z\right),
\eq
where
$$
\phi(u,z)=
\frac
{\vth(u+z)\vth'(0)}
{\vth(u)\vth(z)},
$$
and 
$$
\vth(z|\tau)=q^{\frac{1}{8}}
\sum_{n\in {\bf Z}}(-1)^ne^{\pi i(n(n+1)\tau+2nz)}
$$
is the odd theta-function. Consider the Lax operator
\beq{0d}
L(z)=S_0{\rm Id}+i{\bfS}(z),
\eq
where
$$
\bfS(z)=\sum S_\al\vf_\al(z)\si_\al
$$
and $\si_\al$ are the Hermitian sigma matrices
$\si_\al\si_\be=-i\ep_{\al\be\ga}\si_\ga$.
Then (\ref{0}) is equivalent to the matrix relation
\beq{0e}
\{L_1(z),L_2(w)\}=[r(z-w),L_1\otimes L_2],
\eq
where $r$ is the classical $\SL2$ elliptic r-matrix
\beq{of}
r(z)=\vf_\al(z)\si_\al\otimes\si_\al.
\eq
Note that 
\beq{Cas}
\det (L^2)=K_0\wp(z)-K_1.
\eq

\subsection{Sklyanin algebra from canonical brackets}

Let $E_2$ be the principle $\GL2$-bundle over $\Si_\tau$ with
deg$(E_2)=1$. 
Consider the group element $g\in \Om^{(0)}(\Si_\tau,\GL2)$
$$
g(z,\bz)=g_0(z,\bz)+ic\sum_\al g_\al(z,\bz)\si_\al,
$$
where $c$ is a relativistic parameter. In what follows 
for brevity we omit the $\bz$ dependence.
To be a section of the degree one bundle the functions $g_\al$ should
satisfy the quasi-periodicity conditions 
\beq{9.0a}
g(z+1)=\si_3g(z)\si_3^{-1},
~~g(z+\tau)=\La g(z)\La^{-1},
\eq
$$
~~\La=\La(z;\tau)=-\si_1\exp -\pi(\oh\tau+z).
$$

The holomorphic structure on $E_2$ is defined by
the operator
$$
d_{\bA}=k\bp+\bA:~
\Om^{(j,k)}(\Si_\tau,\sl2)\to \Om^{(j,k+1)}(\Si_\tau,\sl2),
$$
$$
\bA=A_0+i\sum_\al\bA_\al\si_\al.
$$
Here we have introduced the central charge $k$.
The moduli space ${\cal M}$ of stable holomorphic bundles is the factor space
$$
{\cal M}=\{k\bp+\bA\}/{\cal G},
$$
where ${\cal G}=\{f\in \Om^{(0)}(\Si_\tau,\GL2)\}$ is the gauge group
\beq{9.0}
\bA\to f^{-1}k\bp f+f^{-1}\bA f.
\eq

There exists the canonical symplectic structure on the pair of fields\\
${\cal R}'=\{g,\bA\}$
\beq{9.3}
\om'=\int_{\Si_\tau}\tr\left(
D(\bA g^{-1})\we D g\right )
+\frac{k}{2} \int_{\Si_\tau}\left( g^{-1}Dg\we \bp(g^{-1}D g)\right ).
\eq
The canonical class of $\Si_\tau$ is incorporated in the integrals. 
We fix it in what follows.

If $c\to 0$ then in the first order in $c$ we come to the symplectic
structure on the cotangent bundle to the holomorphic structures on 
$E_2$.\cite{Hi}
The element $\sum_\al g_\al\si_\al$ being multiply on the canonical class
plays the role of the Higgs field in the Hitchin construction. 

The form generates the following Poisson brackets in the
space of smooth functionals over ${\cal R}'$
$$
\{H_1(\bA,g),H_2(\bA,g)\}=
$$
$$
\int_{\Si_\tau}\tr\left( \left[\frac {\de H_1(\bA,g)}{\de \bA},
\frac {\de H_2(\bA,g)}{\de \bA}\right]\bA\right) +
k\int_{\Si_\tau}\tr\left(\frac {\de H_1(\bA,g)}{\de \bA}
\bp\frac {\de H_2(\bA,g)}{\de \bA}\right) 
$$
$$
+\int_{\Si_\tau}\tr\left(\frac {\de H_2(\bA,g)}{\de \bA}
\frac {\de H_1(\bA,g)}{\de g}g\right)-
\int_{\Si_\tau}\tr\left(\frac {\de H_1(\bA,g)}{\de \bA}
\frac {\de H_2(\bA,g)}{\de g}g\right).
$$
In particular,
\beqn{9.3b}
\{\bA_\al (z),\bA_\be (w)\}=
\ep_{\al\be\ga}\bA_\ga (z)\de^{(2)}(z-w)+
\frac{k}{2}\de_{\al,\be}\bp\de^{(2)}(z-w),\nonumber\\
\{\bA_0 (z),\bA_0 (w)\}=\frac{k}{2}\de_{\al,\be}\bp\de^{(2)}(z-w),
\eqn
\beqn{9.3e}
\{\bA_0 (z),g_\al (w)\}=\oh g_\al (z)\de^{(2)}(z-w),\nonumber\\
\{\bA_0 (z),g_0 (w)\}=\oh g_0 (z)\de^{(2)}(z-w),\nonumber \\
\{\bA_\al (z),g_\be (w)\}=-\frac{1}{2}
\ep_{\al\be\ga}g_\ga (z)\de^{(2)}(z-w)
-\oh\de_{\al,\be}g_0(z)\de^{(2)}(z-w),\nonumber\\
\{\bA_\al (z),g_0 (w)\}=\oh g_\al (z)\de^{(2)}(z-w),
\eqn
\beq{9.3d}
\{g_\al,g_\be \}=\{g_0,g_\be \}=0,
\eq
where $\de^{(2)}(z-w)$ is the functional 
$f(0)=\int_{\Si_\tau}f(w)\de^{(2)}(z-w)$.

Consider the canonical transformation of the fields
$$
\bA_0\to\bA_0+c(z,\bz),~~\bA_\al\to\bA_\al,~~g\to g,
$$
where $c(z,\bz)\in\Om^{(0,1)}(\Si_\tau)$.
The corresponding moment map is 
 the functional 
\beq{mom}
 \log\det g(z)=\log(g_0^2+\sum_\al g_\al^2).
\eq
Let fix the gauge as $\bA_0=0$.
The form $\om'$ being pushed down on the reduced phase space 
\beq{red}
{\cal R}''=\{g,\bA~|~\bA_0=0,~g_0^2+\sum_\al g_\al^2={\rm const}\}
\eq
is non-degenerate. The Poisson structure on ${\cal R}''$ has the form
(\ref{9.3b}),(\ref{9.3e}),(\ref{9.3d}) where $\bA_0=0$. The determinant
\beq{Ca}
\det g(z)=g_0^2+\sum_\al g_\al^2
\eq
is the Casimir functional with respect to the new brackets.

To complete the model we add some finite degrees of freedom.
Consider the cotangent bundle $T^*\SL2$ represented by
$\bfT\in\sl2,~h\in\SL2$ with the canonical form
\beq{9.5}
\om_0=\tr\left(D(\bfT h^{-1})\we Dh\right).
\eq
The whole upstairs phase space is
$$
{\cal R}_2={\cal R}''\cup T\SL2\sim\{d_{\bA},g,\bfT,h\}
$$ 
with the symplectic structure 
\beq{9.6}
\om=\om'+\int_{\Si_\tau}\om_0\de^{(2)}(z,\bz).
\eq
The form $\om$ is invariant under the gauge transformation (\ref{9.0})
accompanied by
\beq{9.4}
g\to f^{-1}gf,~~(f\in{\cal G}),
\eq
$$
\bfT\to f_0^{-1}\bfT f_0,~~h\to f_0^{-1} h f_0,~~f_0=f(0),
$$
These transformations generate the moment map 
$$
\mu:{\cal R}_2\to{\rm Lie}^*({\cal G})
\mu= -\bA+g\bA g^{-1}-k\bp gg^{-1}-\bfT\de^2(z).
$$
In the case of degree one bundles on elliptic curves
the connection $\bA$ can be 
represented as the pure gauge $\bA=f^{-1}\bp f$. The same gauge transform
brings $g$ in the form
\beq{9.7}
g=f^{-1} L f.
\eq
We preserve the same notations for the transformed variables $\bfT$ and $h$.
Then the moment condition $\mu=0$ takes the form
\beq{9.8}
k\bp L L^{-1}=\bfT \de^2(z,\bz).
\eq
Solutions of this equation should satisfy the consistency condition
\beq{9.9}
k({\rm Res}|_{z=0}L)=\bfT({\rm Res}|_{z=0}L),
\eq
(see \cite{AFM}).
Taking into account the quasi-periodicity of $g$ (\ref{9.0a})
we can write the general solution in the form (\ref{0d})
$$
L(z)=S_0+i\sum_\al S_\al\vf_\al(z)\si_\al.
$$
Then (\ref{9.9}) takes the form
\beq{9.10}
-ikS_\al=\ep_{\al\be\ga}T_\be S_\ga,
\eq
$$
\sum_\al S_\al T_\al=0.
$$
The last equation is consistent with the system (\ref{9.10}).
To come to  nontrivial solutions $\vec{S}$ of  (\ref{9.10}) we 
should assumed that its  determinant vanishes. It is happened
 in the case $k=0$, or 
\beq{9.11}
k^2=\sum_\al T_\al^2.
\eq
This condition restricts the space $T^*\SL2$ to the coadjoint
orbit of $\SL2$. Fortunately, we still have at hand the constant
gauge transformations preserving the gauge condition $\bA=0$. 
The condition (\ref{9.11}) is the moment constraint in
the symplectic reduction of $T^*\SL2$ generating by the constant matrices.
In this way we come to the Lax operator $L$  (\ref{0d}),
where $S_\al$ are related to the elements of the coadjoint orbit
via the solutions of (\ref{9.10}).

Finally, we should demonstrate that the brackets (\ref{9.3b}),
(\ref{9.3e}), (\ref{9.3d}) being restricted on shell
 (\ref{9.8}) after the gauge fixing $\bA=0$ lead
to the Sklyanin algebra (\ref{0}). Using the Dirac procedure we 
define 
\beq{9.12}
\{g(z),g(w)\}^*|_{\rm on~shell}=
-\int_{\Si_\tau}\int_{\Si_\tau}\!\!\!\! dvdu
\{g(z),\bA(v)\}C(v,u)\{g(w),\bA(u)\}|_{\rm on~shell}\;.
\eq
On shell  only
the second term in the r.h.s. of (\ref{9.3b}) is essential.
In this way $C(v,u)$ is the inverse operator to 
$$
k\bp: \Om^{(0)}(\Si_\tau,\sl2)\to \Om^{(0,1)}(\Si_\tau,\sl2),
$$
$$
C(v,u): \eta(u)\in\Om^{(0,1)}(\Si_\tau,\sl2)
\to \int_{\Si_\tau}G(v-u)\eta(v)\in\Om^{(0)}(\Si_\tau,\sl2).
$$
Note, that its action is diagonal
$$ 
\eta(u)_\al\to \int_{\Si_\tau}G_\al (v-u)\eta_\al(v)
$$
Taking into account the quasi-periodicity (\ref{9.0a}) we find 
$$
G_\al=\vf_\al(v).
$$
Substituting in (\ref{9.12}) $g(z)|_{\rm on~shell}=L(z)$
and $C(v-u)$ we come to the Yang-Baxter equation (\ref{0e})
with $r(z-w)$ replaced by $\f1{k}r(z-w)$.
In this way, after the symplectic reduction
of the upstairs space ${\cal R}_2$ we come 
to the two-dimensional phase space ${\cal R}^{\cal S}=\{S_0,\bfS\}/(K_0,K_1)$.

\begin{rem}
The upstairs space ${\cal R}'=\{g,\bA\}$ is similar to the
upstairs space for the
elliptic Ruijsenaars-Schneider (ERS) system.\cite{AFM} 
The only difference is that for ERS system the bundle $E_2$ is trivial 
(deg$(E_2)=0$). There exists the symplectic map from the 
phase space ${\cal R}^{ERS}$ of the two-body ERS to ${\cal R}^{\cal S}$.
In fact, it is the same map that works for the two-body
elliptic Calogero-Moser (ECM) system - $\SL2$-elliptic rotator
correspondence.\cite{LOZ}  It follows from the statement that ERS and ECM are
governed by the same $r$-matrix.\cite{Su} This symplectic map is a twist of the
$r$-matrices. 
 Note that the Hamiltonian
of the elliptic rotator $H^{rot}$ comes from the representation of the 
Casimir
$$
K_1=\oh S_0+H^{rot}.
$$
\end{rem}

\subsection{Symplectic reduction in the Sklyanin algebra}

Here reproduce part of results from  Ref.~\protect\citebk{Do}.
To accomplished our general scheme we introduce the phase space
$$
{\cal R}_{\cal S}=\{S_0,\vec{S};\xi_0,\vec \xi\},
$$
and the symplectic form (\ref{7.2a})
$$
\om=D\xi_0\we DS_0+D\vec \xi\we D\vec S.
$$
Note that we put here the central charge $k=0$ and therefore the 
Poisson sigma-model is one-dimensional.
The form $\om$ is invariant with respect the algebroid action
(\ref{dx}), (\ref{7.5})
$$
\de_{\ve}S_0=-2J_{\be\ga}S_\be S_\ga\ve_\al, ~~ 
\de_{\ve}S_\al=-2\ep_{\al\be\ga}S_0S_\ga\ve_\be,
$$
$$
\de_{\ve}\xi_0=-2\ep_{\al\be\ga}S_\ga\xi_\be\ve_\al
$$
$$
\de_{\ve}\xi_\al=2J_{\al\be}S_\be\xi_\ga\ve_0-
2\ep_{\al\be\ga}S_0\xi_\ga\ve_\be.
$$
The moment map takes the form
$$
\mu_0=2J_{\be\ga}S_\be S_\ga\xi_\al,~~
\mu_\al=2\ep_{\al\be\ga}S_0S_\ga\xi_\be-2J_{\be\ga}S_\be S_\ga\xi_0.
$$
Assume that 
\beq{9.20}
\mu_0=\mu_2=\mu_3=0,~~\mu_1=2\nu.
\eq
The Hamiltonian vector field $V_\ve$ preserving this form of the
moment has the form
$$
\de_\ve S_1=0,~~\de \xi_1=0,
$$
$$
\de_\ve S_0=J_{23}S_2S_3\ve,~~\de_\ve\xi_0=(\xi_2 S_3-\xi_3S_2)\ve,
$$
\beq{9.21}
\de_\ve S_2=-S_0S_3\ve,~~    \de_\ve\xi_2=-(S_0\xi_3+J_{23}S_3\xi_0)\ve,
\eq
$$
\de_\ve=S_0S_2\ve,~~    \de_\ve=(S_0\xi_2-J_{23}S_2\xi_0)\ve.
$$

The reduced phase space 
$$
{\cal R}_{\cal S}^{red}=
\{\mu_0=\mu_2=\mu_3=0,\mu_1=2\nu\}/\de_\ve~{\rm action}~(\ref{9.21})
$$
 is parameterized by the canonical coordinates $(v_1,v_2;u_1,u_2)$
such that
$$
\xi_0=u_1,~~\xi_1=0,~~\xi_2=0,~~\xi_3=u_2,
$$
$$
S_0=v_1,~~S_1=0,~~S_2=-\frac{\nu}{v_1u_2-J_{32}v_2u^1},
~~S_3=v_2,
$$
and
$$
\om|_{{\cal R}_{\cal S}^{red}}=\om^{red}=Dv_1\we Du_1+Dv_2\we Du_2.
$$
In these coordinates the Casimirs (\ref{K}) being reduced on
 ${\cal R}_{\cal S}^{red}$
take the form
$$
K_0=\frac{v_1^2}{2}+\frac{\nu^2}{2(v_1u_2-J_{32}u_1v_2)^2},~~~
K_1=\oh(v_1^2+J_{32}v_2^2).
$$
Since they commute this two-dimensional system is completely
integrable. 
In the trigonometric limit ($Im\tau\to\infty$) 
$e_2=e_3$ and thereby $J_{32}=0$. Then $v_1=const$ and $K_0$ describes
the rational two-body Calogero system with the coupling constant
depending on $v_1$. The Sklyanin algebra is still quadratic
in this limit. Nevertheless, the result of the symplectic reduction 
is essentially the same as when the upstairs space has  the linear 
$\gl2$ brackets.\cite{OP}

\section*{Acknowledgments}
\addcontentsline{toc}{section}{\numberline{}Acknowledgments}
I am grateful for hospitality 
the Max-Planck-Institut f\"{u}r Mathematik (Bonn), where the paper was
 prepared. 
I would like to thank A.~Odessky for illuminating discussion about
the elliptic algebras. The work was partly supported by grants
RFBR-00-02-16530, 00-15-96557 for support of scientific schools,
and INTAS-99-01782.

\section*{References}
\addcontentsline{toc}{section}{\numberline{}References}


\begin{thebibliography}{99}

\bibitem{MM}
M.S.~Marinov and M.V.~Terentev, ``Dynamics on the group manifolds
and the path integral,'' {\em Fortschr.\ Phys.}  {\bf 27}, 511 (1979).
\bibitem{LO}
  A.~Levin and M.~Olshanetsky, 
 ``Hamiltonian Algebroid Symmetries in W-gravity and Poisson sigma-model,''  hep-th/0010043.
\bibitem{Ba}
I.~Batalin, ``Quasigroup construction and first class
constraints,'' {\em  J.\ Math.\  Phys.}  {\bf 22}, 1837 (1981).
\bibitem{HT}
M.~Hennaux and C.~Teitelbom, {\em Quantization of Gauge Systems}, 
(Princeton Univ.\ Press, Princeton, New Jersey, 1992).
\bibitem{I}\ N.~Ikeda, ``Two-dimensional gravity and nonlinear gauge theory,''
{\em Ann.\ Phys.} {\bf 235},  435-464 (1994).
\bibitem{SS}
 P.~Schaller and Th.~Strobl, ``Poisson Structure Induced
(Topological) Field Theories,'' {\em Mod.\ Phys.\ Lett.} A {\bf 9}, 3129-3136 (1994).
\bibitem{P}
 A.~Polyakov, ``Gauge transformations and diffeomorphisms,''
{\em Int. Jorn. Mod. Phys.} A {\bf 5}, 833-842 (1990).
\bibitem{BFK}
 V.~Fock, ``Towards the geometrical sense of operator
expansions for chiral currents and $W$-algebras,'' Preprint ITEP (1990);\\
 A.~Bilal,  V.~Fock and Ia.~Kogan, ``On the origin
of $W$-algebras,''{\em Nucl.\ Phys.} B {\bf 359}, 635--672 (1991).
\bibitem{GLM} 
A.~Gerasimov, A.~Levin and A.~Marshakov,
``On $W$-gravity in two-dimensions,''  {\em Nucl.\ Phys.} B {\bf 360}, 
537-558 (1991).
\bibitem{Ca}
 S.~Carlip, {\em Quantum gravity in 2+1 dimensions,}
(Cambridge Univ. press, Cambridge  1998).
\bibitem{Te}
 C.~Teleman, ``Sur les structures homographiques d'une
surface de Riemann,'' {\em Comment.\ Math.\ Helv.} {\bf 33} , 206-211 (1959).
\bibitem{BD}
 A.~Beilinson and V.~Drinfeld,  ``Opers,''  preprint (1993).
\bibitem{Ad}
 M.~Adler, ``On a trace functional for formal pseudodifferential
operators and the symplectic structure of the Korteveg-de Vries equations,''
{\em Inv. \ Math.} {\bf 50}, 219-248 (1979).
\bibitem{GD}
I.M.~Gelfand and L.A.~Dikii, ``A family of Hamiltonian structures
related to nonlinear integrable differential equations,'' in {\em Collected papers
of I.M.Gelfand,} Vol {\bf 1}, Berlin, Heidelberg, New York : Springer (1987).
\bibitem{BK}
 S.~Barannikov and M.~Kontsevich, ``Frobenius manifolds and
formality of Lie algebras of polyvector fields,''
{\em Internat.\ Math.\ Res.\ Notices} No 4, 201-215  (1998).

\bibitem{TM}
\ J.~Thierry-Mieg, ``BRS analysis of Zamolodchikov's spin 2
and 3 current algebras,'' {\em Phys.\ Lett.} B{\bf 197},  368-372 (1987).
\bibitem{Skl}
 E.~Sklyanin, ``Some algebraic structures connected with
the Yang-Baxter equation,'' {\em Funct.\ Anal.\ Appl.} {\bf 16}, 283  (1982).
\bibitem{AFM}
 G.~Arutyunov, S.~Frolov and P.~Medvedev,
``Elliptic Ruijsenaars-Schneider model from the cotangent bundle over
 the two-dimensional current group,''   hep=th/9608013.
\bibitem{Do}
  V.~Dolgushev,  ``Sklyanin Bracket and Deformation of
 the Calogero-Moser System'', {\em Mod.\ Phys.\ Lett.} A {\bf 16}, 1711 (2001).  
\bibitem{Ma}
 K.~Mackenzie, ``Lie Groupoids and Lie Algebroids
in Differential Geometry,'' {\em London \ Math.\ Soc.\ Lect.\ Notes series,
Cambridge\ Univ.\ Press,} {\bf 124}, (1987).
\bibitem{KM}
 M.~Karasev and V.~Maslov, {\em Nonlinear Poisson Brackets.
Geometry and Quantization,} (in Russian), (Nauka, 1991).
\bibitem{We}
 A.~Cannas~da~Silva and A.~Weinstein, {\em Geometric Models
for Noncommutative Algebras,} (Berkeley Mathematical Lectures,
 Amer.\ Math.\ Soc., Providence {\bf 10}, 1999).
\bibitem{Ka}
 M.~Karasev, ``Analogues of objects of the theory of Lie groups
for nonlinear Poisson brackets,'' {\em Math.\ USSR \ Izvestia}
 {\bf 28}, 497-527 (1987).
\bibitem{Fu}
 B.~Fuchssteiner, ``The Lie algebra structure of degenerate
Hamiltonian and bi-Hamiltonian systems'' {\em Progr.\ Theor.\ Phys.}
 {\bf 68}, 1082-1104 (1982).
\bibitem{H2}
  N.~Hitchin, ``Lie groups and Teichmuller theory,''
{\em Topology} {\bf 31}, 451-487 (1992).
\bibitem{Akh}
 N.I.~Akhiezer, `` Elements of the theory of elliptic functions,''
{\em AMS, Providence}, (1990)
\bibitem{Hi}
 N.~Hitchin, ``Stable bundles and integrable systems'',
{\em Duke \ Math. \ Jour.} {\bf 54}, 91-114 (1987).
\bibitem{LOZ}
  A.~Levin, M.~Olshanetsky and A.~Zotov,
``Hitchin Systems - symplectic maps and two-dimensional version'', 
 nlin.SI/0110045.
\bibitem{Su}
 Yu.~Suris, 
``Elliptic Ruijsenaars-Schneider and Calogero-Moser hierarchies are governed
 by the same r-matrix,'' 
  solv-int/9603011.
\bibitem{OP}
 M.~Olshanetsky and A.~Perelomov, 
 ``Explicit solution of the
Calogero model in the classical case and geodesic flows on symmetric spaces
of zero curvature,'' {\em Lett.\ Nuovo.\ Cim.} {\bf 16}, 333-339 (1976).
\end{thebibliography}
\end{document}